\documentclass[a4paper,12pt]{article}
\usepackage{shadow,pifont,epsfig,wrapfig,times,graphics,color}
\usepackage{latexsym}

\def\beq{\begin{equation}}
\def\eeq{\end{equation}}

\begin{document}

\title{Extra S11 and P13 in the Hypercentral Constituent Quark Model}

\author{M.M. Giannini, E. Santopinto and A. Vassallo\\
{\small Dipartimento di Fisica dell'Universit\`a di Genova
      and I.N.F.N., Sezione di Genova}}

\maketitle

\abstract{We report on the recent results of the hypercentral Constituent
Quark Model (hCQM). The model contains a spin independent three-quark
interaction which is inspired by Lattice QCD calculations and reproduces
the average energy values of the $SU(6)$ multiplets.  The splittings 
within each multiplet are obtained with a $SU(6)$-breaking interaction, 
which can include also an isospin dependent term.\\
All the 3- and 4-stars resonances are well reproduced. Moreover, as all the 
Constituent Quark models, the hCQM predicts ``missing'' resonances 
({\em e.g.} extra $S11$ and $P13$ states) which can be of some help for 
the experimental identification of new resonances.\\
The model provides also a 
good description of the medium $Q^2$-behavior of the electromagnetic 
transition form factors. In particular the calculated helicity amplitude 
$A_{\frac{1}{2}}$ for the $S_{11}(1535)$ resonance agrees very well with
the recent CLAS data.  More recently, the elastic nucleon form factors have 
been calculated using a relativistic version of the hCQM and a relativistic 
quark current.} 

\section{Introduction}
In recent years much attention has been devoted to the description of baryons 
in terms of three quark degrees of freedom. 
Starting from the classical Isgur-Karl model \cite{is}, many different CQMs 
have been developed: the algebraic one
\cite{bil}, the hypercentral CQM \cite{pl} and the GBE model 
\cite{olof,ple}. In the following will be shown the main features of the 
hCQM and will be presented some of the results obtained in the calculation 
of various baryon properties.
\section{The Model}
\label{section:uno}
The internal quark motion is well described by the Jacobi coordinates
{\boldmath $\rho$} and {\boldmath $\lambda$},
or, in an equivalent way, by the hyperspherical coordinates 
\cite{hca}:
\beq
\label{eq:coordinateipercentrali}
x=\sqrt{\rho^2+\lambda^2}\qquad t=\arctan\left(\frac{\rho}{\lambda}\right)
\eeq
\noindent where $x$ is the hyperradius and $t$ is the hyperangle. 
In the hCQM the $SU(6)$-invariant part of the potential is assumed to be 
dependent only on the hyperradius and of the form\cite{pl}: 
$V(x)=-\frac{\tau}{x}+\alpha x\;.$\\
Interaction of the kind linear plus Coulomb-like have been used since time for 
the meson sector ({\em e.g.} the Cornell potential), and has been 
supported by recent Lattice QCD calculations \cite{bali}. 
The choice of an  hypercentral 
potential ({\em i.e.} a potential which depends only on the 
hyperradius) has two different motivations: $x$ is a collective coordinate, 
therefore an hypercentral potential contains also three body effects, 
moreover this potential can be read as the 
hypercentral approximation of a $2$ body potential.\\
The splitting within each multiplet is produced introducing a 
perturbative $SU(6)$-breaking term, which, as a first approximation, 
can be assumed to  
be the standard hyperfine term $H_{hyp}$.\\
The three quark Hamiltonian in the hCQM is then:
\beq
\label{eq:hamiltonianoiperfine}
H = \frac{p_{\lambda}^2}{2m}+\frac{p_{\rho}^2}{2m}-\frac{\tau}{x}~
+~\alpha x+H_{hyp},
\end{equation}
where $m$ is the quark mass (taken equal to $1/3$ of the nucleon mass).
The strength of the Hyperfine interaction is determined in order to reproduce 
the $\Delta$ - $N$ mass difference while the remaining parameters 
($\alpha$ and $\tau$) are fitted to the spectrum, leading to the following 
values: $\alpha=1.61\;\mbox{fm}^{-2}\;,\; \tau=4.59\;.$\\
Keeping these parameters fixed, the resulting wave functions have been used to 
calculate various physical quantities of interest: the helicity amplitudes 
\cite{aie}, 
the electromagnetic transition form factors \cite{aie2}, 
the elastic nucleon form factors \cite{mds}, 
the ratio between electric and magnetic proton form factors \cite{rap}, 
and some interesting quantities related to the parton distributions
\cite{vdhrat}.    
\section{Generalized SU(6)-breaking term}
\begin{figure}[ht]
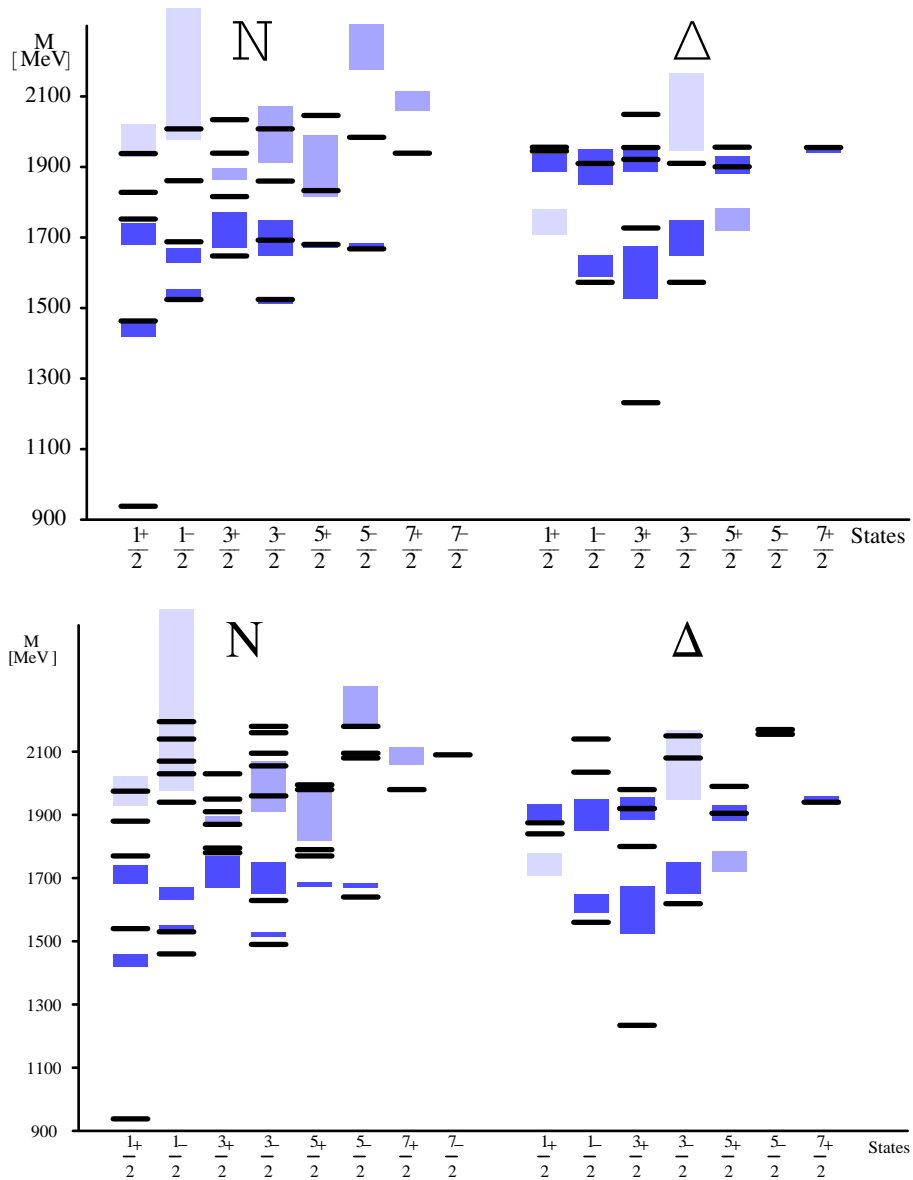

\begin{center}
\includegraphics[width=12cm]{Missing_hCQM.epsi}\vspace*{0.5cm}
\includegraphics[width=12cm]{capstick.fig.epsi}   
\caption{The spectrum obtained with the hypercentral model using the 
interaction in eq.(\ref{eq:interazionecompleta}) (top figure) compared with
the spectrum obtained by Capstick \cite{is} (bottom figure). 
The boxes are the experimental data of PDG with their uncertainties
; dark grey boxes are 3- and 4-stars resonances, 
light grey boxes are 1- and 2- stars resonances\cite{pdg}.\label{fig:spettro}}
\end{center}
\end{figure} 
\noindent
The Hamiltonian of eq.(\ref{eq:hamiltonianoiperfine}) 
give rise to a nice description of the spectrum, nevertheless 
in order to improve the quality of the reproduction, one can  
generalize the Hamiltonian operator introducing 
an isospin dependence in it.\\
The complete interaction used is \cite{iso}:
\beq
\label{eq:interazionecompleta}
H_{int}=V(x)+H_S+H_I+H_{SI}   
\eeq
where $V(x)$ is the $SU(6)$-invariant part, 
$H_S$ is a smeared standard hyperfine 
term, $H_I$ is an isospin dependent term and $H_{SI}$ is a spin-isospin 
dependent term. The spectrum obtained with the interaction of eq.
(\ref{eq:interazionecompleta}) is shown, in fig.\ref{fig:spettro}.
All the 3- and 4-stars resonances, in particular the Roper, are well 
reproduced.\\
All the CQMs predict states which don't have (yet) experimental 
confirmsations. In particular (see Table \ref{tab:missing} and 
Fig.\ref{fig:spettro}) the hCQM predicts 5 missing resonances with energies 
below $1900$ MeV. Recent analysis (see for example \cite{saghai,ripani} and 
references quoted therein) show that there are 
some indications for the presence of a third $S11$ and a third $P13$ 
with masses comparable with the predictions of the hCQM.

\begin{table}[ht]
\caption{hCQM prediction for S11, D13, P13 and P33 resonances, compared with 
PDG data\cite{pdg}. \vspace*{1pt}}{ 
\begin{tabular}{|c|r|r|r|r|r|r|r|}
\hline
State & PDG & hCQM & hCQM+Iso & State & PDG    & hCQM & hCQM+Iso\\
\hline
{\bf S11} &  1535  & 1507 & 1524    &   {\bf P13} &  1720  & 1797 & 1848    \\
    &  1650  & 1574 & 1688    &       &  1900  & 1835 & 1816    \\
    &        & 1887 & 1861    &       &        & 1853 & 1894    \\
    & (2090) & 1937 & 2008    &       &        & 1863 & 1939    \\
\hline
{\bf D13} &  1520  & 1526 & 1524    &   {\bf P33} &  1232  & 1240 & 1232    \\ 
    &  1700  & 1606 & 1692    &       &  1600  & 1727 & 1723    \\
    &        & 1899 & 1860    &       &  1920  & 1843 & 1921    \\
    & (2080) & 1969 & 2008    &       &        & 1856 & 1955    \\
    &        &      &         &       &        & 2104 & 2049     \\    
\hline
\end{tabular} \label{tab:missing}}
\end{table}
\section{The electromagnetic transition form factors}
\begin{figure}[t]
\begin{center}
\centerline{\epsfxsize=4.5in\epsfbox{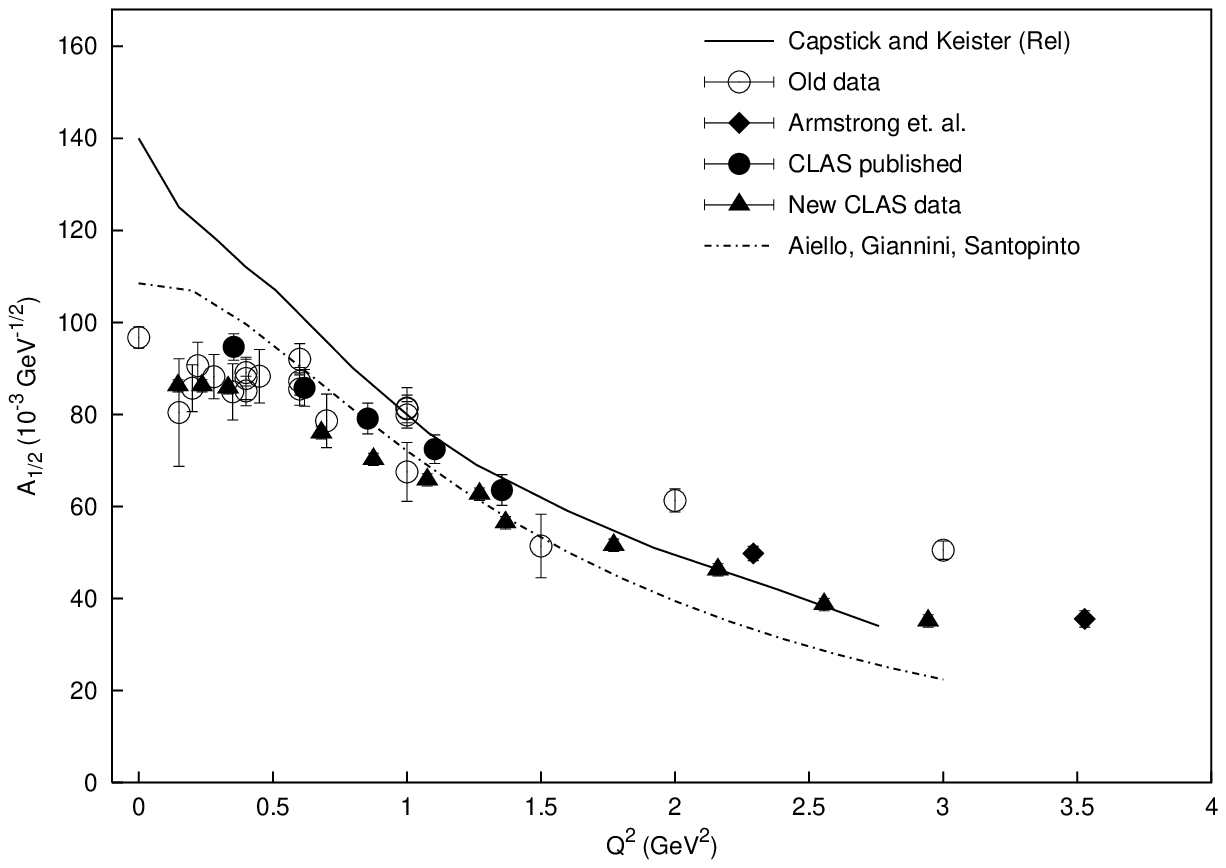}}   
\caption{The helicity amplitude $A^p_{1/2}$ for the $S11(1535)$ resonance 
calculated with the hCQM (dashed curve) and with the model of Ref. 
\cite{capstick} (full curve). The data are taken from the compilation of 
Ref. \cite{burk}.\label{fig:s11}}\vspace*{-0.4truecm}
\end{center}
\end{figure}

The helicity amplitudes for the e.m. excitation of baryon resonances, 
$A_{1/2}$, $A_{3/2}$ and $S_{1/2}$ are calculated as the transition matrix  
element of the transverse and longitudinal part of the e.m. 
interaction between the nucleon and the resonance states given by this model. 
A non relativistic current for point quarks has been used.\\
The longitudinal and transverse transition form factors have been 
systematically calculated for all the resonances (including the missing ones) 
predicted by the hCQM. The results for the $A^p_{1/2}$ and $A^p_{3/2}$ 
amplitudes 
for all the negative parity resonances are reported in Ref. \cite{aie2}. In 
fig.\ref{fig:s11} the result for the $A_{1/2}$ amplitude for the 
$S11(1535)$ is shown; the prediction agrees quite 
well with the data except for some discrepancies at small $Q^2$. These 
discepancies could be ascribed to the non-relativistic character of the model, 
or better to the lack of explicit quark-antiquark 
configurations which are expected  to be important at low $Q^2$.   
\section{Relativity and elastic nucleon form factors}
It is well understood that in order to obtain a better description of the 
baryon properties one has to introduce relativity in CQMs.  
Starting from a CQM one can introduce relativistic effects by: a) using 
a relativistic kinetic energy operator; b) boosting the baryon wave functions 
from the initial and final rest frames to a common frame; c) using a 
relativistic quark current. In the hCQM the potential parameters have been 
refitted using a relativistic kinetic energy operator, the resulting spectrum 
is not much different from the non  relativistic one. The boosts and a 
relativistic quark current expanded up to the lowest order in quark momenta 
has been used both for the elastic form factors\cite{mds} and for the 
helicity amplitudes \cite{mds2}. While in the latter case the effect of these 
relativistic corrections is small, for the elastic form factors the 
relativistic effects are more important, in particular, as we have 
shown for the first time in \cite{rap}, they are responsible 
for the decreasing $Q^2$-behaviour of the ratio between the electric and 
magnetic proton form factors.\\
More recently a relativistic quark current with no expansion in the quark 
momenta and the boosts to the Breit frame have been applied to the calculation 
of the elastic form factors. The resulting theoretical curves 
\cite{unpubblished}, 
calculated without free parameters and with pointlike quarks, 
agree very nicely with the 
experimetal data except for some discrepancies at low $Q^2$. The decrease of 
the ratio between electric and magnetic proton form factors is stronger than 
in the previous cases and reaches almost the $50\%$ level, not far from the 
recent TJNAF data \cite{jones}.

\end{document}